\documentclass[9pt,twocolumn,twoside]{osajnl}

\journal{ol} 

\setboolean{shortarticle}{true}

\usepackage{amssymb,amsmath}
\usepackage{graphicx}
\usepackage{eurosym}
\usepackage{color}
\usepackage{todonotes}
\usepackage[normalem]{ulem}
\usepackage[makeroom]{cancel}

\newcommand{\PT}{\mathcal{PT}}

\newcommand{\rR}{{\rm R}}
\newcommand{\rL}{{\rm L}}
\newcommand{\rX}{{\rm X}}

\newcommand{\mR}{$\mu$R\ }

\newcommand{\rC}{{{\rm C}}}

\newcommand{\rM}{{{\rm M}}}

\newcommand{\rtext}[1]{\textcolor{black}{{#1}}}

\title{
Spectral singularities of a potential created
by two coupled microring resonators}

\author[1,*]{Vladimir V. Konotop}
\author[2]{Barry C. Sanders}
\author[3]{Dmitry A. Zezyulin}

\affil[1]{Centro de Física Teórica e Computacional and Departamento de Física, Faculdade de Ciências, Universidade de Lisboa, Campo Grande, Ed. C8, Lisboa 1749-016, Portugal}
\affil[2]{%
Institute for Quantum Science and Technology,
University of Calgary, Calgary, Alberta T2N 1N4, Canada}
\affil[3]{ITMO University, St.~Petersburg 197101, Russia}

\affil[*]{Corresponding author: vvkonotop@fc.ul.pt}

\begin{abstract}
Two microring resonators, one with gain and one with loss, coupled to each other and to a bus waveguide, create an effective non-Hermitian potential for light propagating in the waveguide. Due to geometry,
coupling for each microring resonator yields two counter-propagating modes with equal frequencies. We show that such a system enables implementation of many types of scattering peculiarities, which are either the second or fourth order. The spectral singularities separate parameter regions where the spectrum is either pure real or else comprises complex eigenvalues; hence, they represent the points of the phase transition. By modifying the gain-loss relation for the resonators such   an optical scatterer can act as a laser, as a coherent perfect absorber, be unidirectionally reflectionless or transparent, and support bound states either growing or decaying in time. These characteristics are observed for a discrete series of  the incident-radiation wavelengths.  

\end{abstract}

\setboolean{displaycopyright}{true}

\begin{document}

\maketitle

Microcavities are multi-functional optical elements employed as element of numerous devices~\cite{review1,Vahala,review2}. In the simplest case of a microring resonator side-connected to a bus waveguide~\cite{Yariv,CaPaVa2000},  amplitude and phase of the field transmitted through
the coupling region are modified~\cite{Heebner}, in particular, allowing for lasing if the microring is filled with active material~\rtext{\cite{Feng2014,Longhi}}.  With two microrings (or microdiscs) connected to a bus waveguide, one can significantly modify  asymmetric spectra of the Fano resonance for transmitted radiation~\cite{three_cav}. More sophisticated effects on a signal transmitted through a bus  waveguide can be created by coupling it to many microrings. In particular, a periodic array of microrings strongly modifies waveguide dispersion~\cite{Heebner}, whereas microcavities assembled in photonic molecules enables incorporating prescribed spectral characteristics of the transmitted light~\cite{phot_molec}.  New functionalities can be achieved for microring or microdisc cavities connected to two waveguides. Micro-cavities in such configurations allow for observing nonreciprocal light propagation~\cite{SoCaAl,nonrecip1,nonrecip2} or revival of lasing by induced losses~\cite{reverse}. 

Generically, light propagating in a waveguide with connected microcavites experiences an effective optical potential, which is non-Hermitian if the cavities are active or absorbing media. Such potentials could possess spectral singularities (SSs), which in optical systems produces
lasing~\cite{Mostafazadeh2013} and coherent perfect absorption (CPA)~\cite{Stone},
or in special cases, CPA-lasing~\cite{Longhi-laser} occurring at the same wavelength. These phenomena were observed in waveguides connected with cavities in parity-time ($\PT$)-symmetric configuration. In~\cite{ReGu} it was reported on possibility of control of SS by cavity parameters.
A microring resonator with  a complex grating respecting $\PT$-symmetry, coupled to a waveguide, allows for single-mode emission and simultaneous coherent perfect absorption~\cite{Longhi}. 

\begin{figure}
	\centering
	\includegraphics[width=0.95\columnwidth]{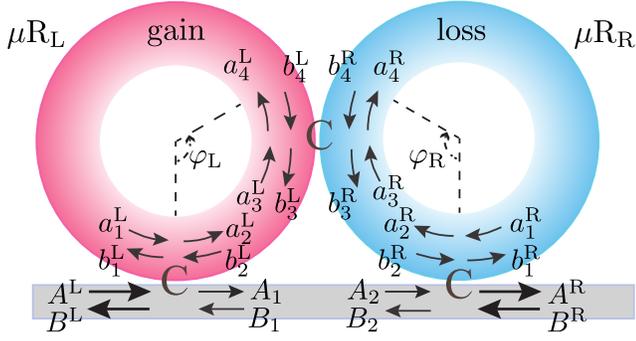}
	\caption{Schematic 
		 of two $\mu$Rs coupled to a bus waveguide (fiber) shown by the gray strip. The left (L) and right (R) $\mu$Rs have respectively amplifying and absorbing media and are characterized by the round trip factors $\alpha_\rL$ and $\alpha_\rR$. The couplers are indicated by  $\rm C$ symbols. Arrows show the fields circulating in the  $\mu$Rs and transformed in the couplers regions. The field amplitudes are indicated next to the arrows. The angular coordinates $\varphi_\rL$ and $\varphi_\rR$ are positive when rotated respectively in counterclockwise and clockwise directions.}
	\label{fig:one}
\end{figure}

Most of these systems deal with microresonators operating in a single-mode regime. To create a non-Hermitian optical potential for a waveguide coupled to microresonators,   at least a two-mode configuration providing forward and backward propagating fields is required. This condition was achieved by using a $\PT$-symmetric grating resulting in coupling two counter-propagating modes of a single microring ($\mu$R)~\cite{Longhi}. In this Letter we describe a device for which coupling of modes in a system comprising two $\mu$Rs, \rtext{similar to ones used in the experiment~\cite{Christod},} connected   to a waveguide occurs due to geometric factors, as  shown schematically in Fig.~\ref{fig:one}. One of the $\mu$Rs is active, assuring gain of the propagating field, and another one is absorbing. Due to the arrangement, each of the connections (in Fig.~\ref{fig:one} they are denoted by C) among the $\mu$Rs and the waveguide   is working in the bidirectional regime, i.e. each of the $\mu$Rs bears two modes circulating in opposite directions while the waveguide carries incident and transmitted waves. The system is free from a quite demanding constraint to obey $\PT$ symmetry (i.e. to have exact balance of gain and losses in the coupled $\mu$R). 
    
Given these simple features, we show that, by changing gain (or loss, or both), one can continuously shift spectral singularities, thereby achieving many of the well known regimes, including  lasing, CPA, CPA-lasing, and bound states which either grow or decay. Moreover, the transfer matrix of the optical potential created by the microrings having identical geometry is characterized by the zeros of the second order, thus, giving origin to \emph{ SSs of second and fourth orders}. \rtext{Such high-order SSs could
have practical importance.
For example, if a small-amplitude wavepacket is incident at a non-Hermitian potential featuring a high-order SS, the   amplitudes of the transmitted and reflected wavepackets increase with the order of the SS~\cite{doubleSS}.
Also high-order SSs allow for controlled creation of simple SSs at different frequencies, which could
be achieved by variation of parameters of the potential~\cite{doubleSS}.}
 
To quantify our system parameters, we assume that the right (denoted by ``R'') absorbing {\mR} is characterized by the round-trip factor, determining the decay of the field amplitude after one round trip~\cite{CaPaVa2000},  $\alpha_\rR\leq 1$. The left (denoted by ``L'') ring with active medium is characterized by the round-trip factor $\alpha_\rL\geq 1$, which characterizes increasing the field amplitude over one round trip. Thus, $\alpha_{\rR},1/\alpha_\rL\in(0,1]$.  Larger values of $\alpha_{\rL}$ correspond to stronger gain, and smaller values of $\alpha_\rR$ correspond to stronger losses; equality of any of the round-trip factors to one corresponds to the conservative microring. 

Due to the geometry in each ring,
two modes propagate. These modes are denoted $a^\rX$ and $b^\rX$, with $\rX\in\{\rL,\rR\}$.
In each $\mu$R the fields $a^\rX$ and $b^\rX$ propagate in opposite directions: $a^\rX\propto e^{ik_\rX\rho\varphi_\rX}$ and  $b^\rX\propto e^{-ik_\rX\rho \varphi_\rX}$, where $k_\rX=2\pi/\lambda_\rX=n_\rX\omega/c$, $\rho$ is the radius of both rings, $\lambda_\rX$ is the wavelength in the respective ring (at this stage the rings could admittedly have different refractive indexes $n_X$), and without loss of generality we consider the positive angular directions to be counterclockwise for $\varphi_\text{L}$ and clockwise for $\varphi_\text{R}$
with $\varphi_{\rL,\rR}\in[0,2\pi)$.
Field amplitudes incident on and transmitted by each of the couplers are distinguished by the lower indexes $j=1,\dots,4$ as explained in Fig.~\ref{fig:one}.
Thus, with our notation, we have the amplitude relations:
\begin{eqnarray}
\begin{array}{ll}
a_1^{\rX}=\phi_\rX^{3/4}\alpha_\rX^{3/4} a_4^X, & a_3^{\rX}=\phi_\rX^{1/4}\alpha_\rX^{1/4} a_2^X, 
\\
b_1^{\rX}=\phi_\rX^{-3/4}\alpha_\rX^{-3/4} b_4^X, & b_3^{\rX}=\phi_\rX^{^-1/4}\alpha_\rX^{-1/4} b_2^X, 
\end{array}
\end{eqnarray}
where, for the sake of convenience, we introduced the phase factors $\phi_\rX=\exp(i2\pi k_\rX\rho)$.
The factors $\phi_{\rL,\rR}$
are related by
$\phi_\rL=\phi_\rR \exp[2\pi i\rho(k_\rR-k_\rL)]$. Thus, using the simplified notation $\phi\equiv\phi_\rL$ below we consider only $\phi$, i.e., the phase of the left \mR,
as a frequency-related variable. 
The couplers connect the fields between the $\mu$Rs ($\top$ stands for transpose):  $\left(a_4^{\rL},a_4^{\rR}\right)^\top=\rC\left(a_3^{\rL}, a_3^{\rR}\right)^\top$, $\left(b_3^{\rL}, b_3^{\rR}\right)^\top=\rC \left(b_4^{\rL},b_4^{\rR}\right)^\top$ for
a lossless coupler, which is expressed by the unimodular property of the matrix~\cite{Yariv,CaPaVa2000}: 
\begin{equation}
\rC=\left(\!\begin{array}{cc}
\tau  & i\kappa \\ i\kappa & \tau
\end{array}\!\!\right), \quad \tau^2+\kappa^2=1
\end{equation}
 (for simplicity, $\tau$ and $\kappa$ are considered real).
Denoting the right (left) propagating beams in the bus waveguide by $A$ ($B$), as shown in Fig.~\ref{fig:one}, analogous equations are valid for coupling between the $\mu$Rs and the bus waveguide: 
$\left(a_2^{\rL},A_1\right)^\top=\rC\left(a_1^{\rL},A^{\rL}\right)^\top$,
$\left(b_1^{\rL},B^{\rL}\right)^\top=\rC\left(b_2^{\rL}, B_1\right)^\top$,
$\left(a_2^{\rR}, B_2\right)^\top=\rC\left(a_1^{\rR}, B^{\rR}\right)^\top$, and
$\left(b_1^{\rR}, A^{\rR}\right)^\top=\rC\left(b_2^{\rR}, A_2\right)^\top$.
Finally, here we consider the conservative waveguide, such that the connection of the fields between the couplers is a pure phase: $A_2=e^{i\theta}A_1$ and $B_2=e^{-i\theta}B_1$.

Below we concentrate on stationary propagation. The $\mu$Rs create an effective non-Hermitian potential for the light propagating in the bus waveguide. The transfer matrix $\rM$ of this potential relates inward and outward propagating waves, from both sides of the region where the   $\mu$Rs are connected to the bus waveguide, i.e.,   $( A^{\rR},B^{\rR})^\top=\rM(\phi)( A^{\rL}, B^{\rL})^\top$.
The matrix $\rM(\phi)$ can be computed in explicit form using a computer algebra program. As its entries are quite bulky, we do not present them in the text. Instead,  below we concentrate specific characteristics of the scattering data.  In particular, we verify that $\det \rM=1$, i.e., reciprocity is not violated. The left and right reflection coefficients are defined as $r_\rL=-M_{21}/M_{22}$ and $r_\rR=M_{12}/M_{22}$, while the two transmission coefficients are equal and given by $t_\rL=t_\rR=t=1/M_{22}$ ($M_{ij}$ with $i,j\in\{1,2\}$ are the entries of $\rM$). 

\rtext{Universality of the device illustrated in Fig.~\ref{fig:one}, which allows different regimes by changing gain or loss parameters, 
is characterized by the location of zeros of the matrix-element $M_{22}$ in the complex plane. Thus, to illustrate the mentioned regimes and to describe  transitions among them, we proceed with the investigation of ``motion''  of the zeros of $M_{22}$ under changes of the system parameters.}

Let~$\phi_j$ be a zero  of $M_{jj}$,
$j\in\{1,2\}$, i.e., $M_{jj}(\phi_j)=0$.
Then, in terms of $\phi_j$, the unbroken phase, i.e., pure transition and reflection, corresponds to all zeros $\phi_1$ inside ($|\phi_1|<1$) and all zeros $\phi_2$ outside ($|\phi_2|>1$) the unit circle of the complex variable $\phi_j$.
\rtext{In both these cases,
the spectrum of the waves in the waveguide is purely real.}
CPA and coherent lasing regimes take place if $|\phi_{1,2}|=1$, respectively.
If either  $|\phi_1|>1$ or $|\phi_2|<1$, then the system supports \rtext{non-stationary bound states. Indeed, propagation of a given mode in the waveguide outside the coupling region is described by the wave equation.
If, at a given frequency, $\omega$, and wavenumber, $k$, one has $M_{22}=0$ (the case $M_{11}=0$ is analyzed similarly), the fields to the right ("$+$") and to the left ("$-$") from the coupling region is obtained as $E_\pm\exp[i(\pm kx-\omega t)]$,  where $x$ is the coordinate along the waveguide and $E_{\pm}$ are the field amplitudes. 
As~$|\phi_2|<1$ corresponds to Im$\,k>0$ and hence Im$\,\omega>0$, the respective solution decays at $|x|\to\pm\infty$, i.e., represents   a bound state with  amplitude growing in time.}

We start with the case that refractive indexes~$n_\text{L}$ and and~$n_\text{R}$ in left and right $\mu$Rs are chosen such that $\rho(k_\text{R} - k_\text{L})=m$, with~$m$ an integer,
which implies that the phase factors are equal: $\phi_\text{L} = \phi_\text{R}=\phi$.  
For~$\alpha$ a ``reduced'' round-trip factor,
the roots $\phi_{1,2}$ are
\begin{equation}
\label{roots}
\frac{1}{\tau}\phi_1^{\pm}=\tau\phi_2^{\pm}=\frac{\tau}{\alpha}\pm\sqrt{\frac{\tau^2}{\alpha^2}-\frac{1}{\alpha_\rL\alpha_{\rR}}},\;
\alpha=2\frac{\alpha_\rL\alpha_\rR}
    {\alpha_\rL+\alpha_\rR}.
\end{equation}
Importantly, these roots are not valid in the case $\alpha_\rL=\alpha_{\rR}=\tau=1$
(see discussion below). We also emphasize that the roots $\phi_{1,2}^{\pm}$ and, hence, the spectral singularities discussed below, are at least of \emph{second order}, i.e., $M_{jj}(\phi_j^{\pm})=[dM_{jj}(\phi)/d\phi]_{\phi_j^{\pm}}=0$. At $\tau=\alpha/\sqrt{\alpha_{\rR}\alpha_\rL}$ the expression under the radical in (\ref{roots}) vanishes and  both zeros merge forming a zero of \emph{fourth order}.
 
The phase transition (i.e., passage of either lasing or absorbing thresholds) is not directly related to the exact balance of gain and absorption: even for imbalanced gain and absorption the phase can be unbroken. The transition can occur according to different scenarios, similar to one described in~\cite{KonZez2017}. We use $\tau^2$ as the control parameter that characterizes the coupling: $\tau^2=1-\kappa^2$: smaller $\tau^2$ corresponds to stronger coupling. It follows from (\ref{roots}) that, in the limit $\tau\to 0$, the phase factors $\phi_{1}^\pm$ are situated at the origin of the complex plane whereas the phase factors $\phi_{2}^\pm$ have indefinitely large amplitudes.

\begin{figure}
	\centering
	\includegraphics[width=\columnwidth]{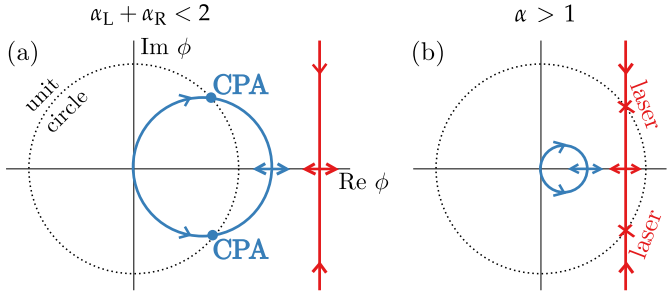}
	\caption{Two scenarios of the phase transition with complex conjugate phase factors in the complex plain $\phi$.  Blue solid circle and red line show the trajectories of $\phi_{1}^\pm$ and  $\phi_{2}^\pm$, respectively. The arrows' directions correspond to increasing $\tau^2$.  (a) If $\alpha_\rL+\alpha_\rR<2$ the   phase transition trough the  CPA  occurs when blue solid semicircles cross (blue filled circles) the unit circle (the dashed line). For increasing $\tau^2$,  the phase factors leave the unit circle,  eventually colliding at the real axis and then split off in opposite directions shown by arrows. (b)  At $\alpha>1$ the phase transition happens through the lasing when the vertical red line crosses (red crosses) the unit circle.}
	\label{scenarios}
\end{figure}

For increasing $\tau<\alpha/\sqrt{\alpha_\rL\alpha_\rR}$,
the phase factors $\phi_{1}^\pm$ move along the opposite (upper and lower) arcs of a circle  centered  at the complex-plane point $((\alpha_{\rR}+\alpha_\rL)^{-1}, 0)$ and having  radius equal to $(\alpha_{\rR}+\alpha_\rL)^{-1}$ [blue circles in Fig.~\ref{scenarios}]. Thus, if $\alpha_{\rR}+\alpha_\rL<2$, then the arcs cross the unit circle at  $\tau^2=\tau^2_{\rm CPA}=\alpha_\rL\alpha_\rR$ [Fig.~\ref{scenarios}(a)]. Notice that the requirement  $\alpha_{\rR}+\alpha_\rL<2$ implies   $\alpha_\rL\alpha_\rR<1$, which means that this CPA scenario occurs only if absorption exceeds gain. At $\tau^2 = \tau^2_{\rm CPA}$ simultaneously two second order CPAs take place for frequencies having equal positive and negative detunings from $\omega_0=cm/(\rho n)$ where $n=n_\text{L}=n_\text{R}$: 
$\omega-\omega_0=\pm (c/2\pi \rho n)\arccos[(\alpha_\rL+\alpha_\rR)/2]$.
For the case $\alpha_{\rR}+\alpha_\rL=2$, two  complex-conjugate phase factors $\phi_{1}^\pm$  collide at the complex-plane point $(1,0)$ and then move along the real axis in opposite directions.
This is the case when the phase transition occurs through a fourth-order spectral singularity. 

When  $\tau^2$  increases starting from zero, the phase factors $\phi_2^\pm$ move towards each other from infinity along the vertical line $\textrm{Re}\,\phi =1/ \alpha$ in the complex plane [red lines in Fig.~\ref{scenarios}]. Thus, if $\alpha>1$, then lasing occurs when the vertical line crosses the unit  circle at $\tau^2 = \tau^2_{\rm las}=1/(\alpha_\rL\alpha_\rR)$ [Fig.~\ref{scenarios}(b)]. The condition  $\alpha>1$ implies $\alpha_\rL\alpha_{\rR}>1$; i.e., gain is dominating. Thus, lasing occurs at frequency detunings $  \omega-\omega_0=\pm (c/2\pi \rho n)\arccos\left[\alpha^{-1}\right]$.
If $\alpha=1$ holds, then the vertical line $\textrm{Re}\,\phi = 1/\alpha$  touches the unit circle at the point  $(1,0)$, and fourth-order phase transition occurs. 

Simultaneous touching of the blue circle with a red straight line at $(1,0)$; i.e. a CPA-laser of the fourth order, is impossible. This stems from the fact that the limiting transitions $r=1$ and $\alpha_{\rL,\rR}\to 1$ (the conservative limit of the disconnected $\mu$Rs) and $\tau\to 1$ and $\alpha_{\rL,\rR}=1$  (the limit of vanishing coupling $\kappa$ of conservative $\mu$Rs) are non commutative.  In the last case we deal with a conservative system that cannot have zeros on the unit circle (in this case Eq.~(\ref{roots}) is not valid). If however,   $\alpha_{\rL}\neq 1$, the decoupling of the system elements leads to infinite growth of the field amplitude at $\tau=1$ (say, $a^L_j\propto \kappa/(\alpha_\rL-1)$ at $\alpha_{\rL}\to 1$).
\begin{figure}
	\centering
	\includegraphics[width=\columnwidth]{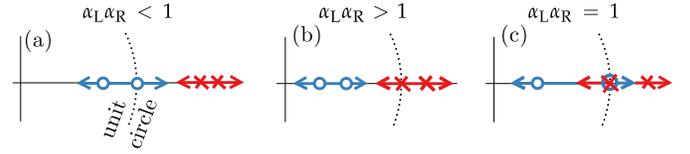}
	\caption{Scenarios of phase transition with  real phase factors. Blue circles and  red crosses show $\phi_{1}^\pm$ and $\phi_2^\pm$, respectively, and the arrows show the directions of the phase factors as $\tau^2$ increases.   (a) If $\alpha_\rL\alpha_\rR<1$, then the phase factor $\phi_1^+$ crosses the unit circle (shown with dashed line) at the point $(1,0)$, and the phase transition through the CPA occurs.   (b)  If $\alpha_\rL\alpha_\rR>1$, then   the phase factor $\phi_2^-$  enters the unit circle, and a phase transition through lasing occurs. (c) In the $\PT$-symmetric configuration with   $\alpha_\rL\alpha_\rR=1$,   phase factors $\phi_1^+$ and $\phi_2^-$ cross the unit circle simultaneously, and  the phase transition happens through a CPA-laser.}
	\label{scenarios_Re}
\end{figure} 

If  $\tau^2>\alpha^2/(\alpha_\rL\alpha_\rR)$, then all four phase factors are positive and move along the real axis for increasing $\tau^2$. The phase transition occurs if one of the phase factors crosses the unit circle at the point $(1,0)$ of the complex plane. If losses are stronger than gain, then the phase transition occurs through the CPA [Fig.~\ref{scenarios_Re}(a)]. In the opposite situation, when gain dominates, then the phase transition corresponds to lasing [Fig.~\ref{scenarios_Re}(b)]. CPA and lasing at the real axis take place, respectively, at
\begin{equation}
\tau^2_{\rm CPA} = \frac{\alpha_\rL\alpha_\rR}{\alpha_\rL+\alpha_{\rR}-1}, \qquad  \tau^2_{\rm las} = \frac{1}{\alpha_\rL+\alpha_{\rR}-\alpha_\rL\alpha_{\rR}}.
\end{equation}
For $\PT$ symmetry, when the strengths of gain and loss are equal, i.e., $\alpha_{\rL}\alpha_\rR=1$, two phase factors  cross the unit circle simultaneously, and the phase transition occurs through the CPA-laser of the \emph{second order} [Fig.~\ref{scenarios_Re}(c)].

\begin{figure}
 	\centering
 	\includegraphics[width=1.0\columnwidth]{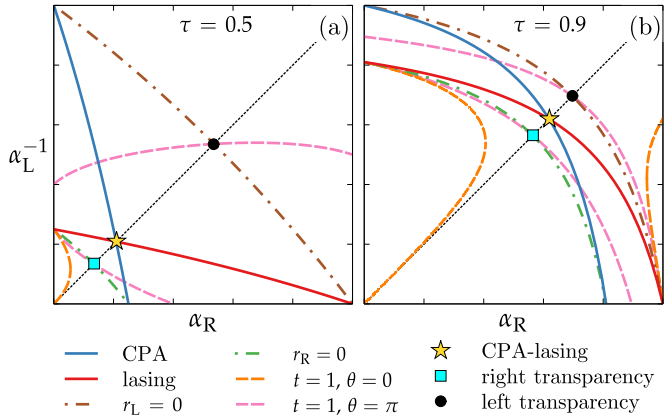} 
 	\caption{Gain and loss parameters
 	at which the effective potential due to the coupled $\mu$Rs posses specific properties: CPA (blue solid curves), lasing (red solid curves), left-- and right-reflectionless propagation (brown and green  dash-dotted lines), and transparency ($t=1$) obtained for $\theta=0$  and $\theta=\pi$ (orange and pink  dashed lines). The results are obtained for the coupling parameter (a) $\tau=0.5$  and (b) $\tau=0.9$. The star indicates the parameters at which CPA-lasing occurs, while square and circle  indicate perfect transparency (reflection zero and transmission one) for the left-- and right-- propagating waves. The thin dotted diagonal lines correspond to the $\PT$-symmetric case $\alpha_{\rR}\alpha_\rL=1$.  The plots are shown in windows $(\alpha_\rR, \alpha_\rL^{-1})\in[0,1]\times[0,1]$.  
 	}
 	\label{fig:two}
 \end{figure}

Using the round-trip factor of the cavity with absorption, $\alpha_\rR$, as a control parameter we compute the round-trip factor of the cavity with gain giving CPA as $\alpha_\rL=\tau^2(1-\alpha_\rR)/(\tau^2-\alpha_\rR)$ and giving laser as  $\alpha_\rL=(1-\tau^2\alpha_\rR)/ [\tau^2(1-\alpha_\rR)]$. These two curves are plotted with solid line  in Fig.~\ref{fig:two} on the plane $(\alpha_\rR,1/\alpha_\rL)$ for two different coupling matrices. We observe, that the increase of coupling (decrease of $\tau$), requires increase of gain and loss for achieving CPA or lasing.
Quite counter-intuitively,
for any fixed dissipation ($\alpha_\rR$),
gain exists for which lasing occurs,
but the converse is generally untrue:
not every value of gain yields lasing. {\it Vice versa}, for any fixed gain ($1/\alpha_\rL$) dissipation exists at which CPA  occurs, but, in order to achieve CPA,
dissipation must be strong enough (i.e., $\alpha_{\rR}$ must be small enough).
  
The effective potential created by the $\mu$Rs displays other interesting characteristics, which are summarized in Fig.~\ref{fig:two}. For any value of gain or loss,
the effective potential can be made left-reflectionless, $r_\text{L}=0$ (brown dash-dotted lines).
For limited intervals of gain or loss, it can be made right-reflectionless, $r_\text{R}=0$ (green dash-dotted lines). These results do not depend on the specific value of the phase~$\theta$ acquired during propagation in the waveguide between the couplers. However, as scattering data themselves depend on the phase gain in the waveguide, the possibility to obtain transparency, i.e., to satisfy the conditions $t=1$, does depend on~$\theta$ (see dashed orange and dashed pink lines for $\theta=0$ and $\theta=\pi$, respectively). The  device is left- and right-transparent with no reflection  only at $\theta=\pi$ and in the $\PT$-symmetric case $\alpha_{\rL}\alpha_{\rR}=1$  (shown by filled dot and squares).

\begin{figure}
	\centering
	\includegraphics[width=\columnwidth]{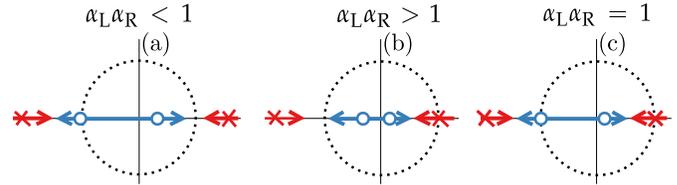}
	\caption{Scenarios of phase transition when the phase factors  $\phi_\rL$ and $\phi_\rR$ in   left and right $\mu$Rs are of opposite sign. (a) Phase transition through the CPA. (b) Phase transition through lasing. (c) Phase transition through the simultaneous CPA and lasing which occur at different frequencies $\omega$. }
	\label{scenarios01}
\end{figure}

Now we briefly discuss a situation for which $\rho(k_\text{R}-k_\text{L})$ is half-integer. In this case phase factors in left and right rings are of opposite signs: $\phi_\rR=-\phi_\rL$. As above, we  use $\phi \equiv \phi_\rL$ as a  parameter. In this case all four phase factors $\phi_{1}^\pm$ and    $\phi_{2}^\pm$ are real and move along the real axis for increasing $\tau^2$. At $\tau=0$ phase factors $\phi_{1}^\pm$ are situated at the origin, and $\phi_{2}^\pm=\pm\infty$. 
Increasing $\tau^2$ makes $\phi_{1}^\pm$ split off in opposite directions, and $\phi_2^\pm$ move towards each other; see Fig.~\ref{scenarios01}.
The phase transition occurs when one of the phase factors crosses the unit circle at point $(1,0)$ or $(-1, 0)$. Depending on the ratio between gain and losses, three different scenarios are possible. If $\alpha_\rL\alpha_{\rR}<1$, i.e., losses exceeds gains, so the phase transition occurs through the CPA, when the phase factor $\phi_1^-$ reaches the point $(-1, 0)$; see Fig.~\ref{scenarios01}(a).  The corresponding value of $\tau^2$ and the frequency amounts to 
$
\tau^2_{\rm CPA}  = 
{\alpha_{\rR}\alpha_\rL}/(1+\alpha_\rL-\alpha_\rR), 
$
$
\omega_{\rm CPA}=c
(m+1/2)/(\rho n_\text{L}).
$
If gain dominates,
the phase transition occurs when the phase factor $\phi_2^+$ reaches the point $(1,0)$; see Fig.~\ref{scenarios01}(b). This happens at
$
\tau^2_{\rm las}  = {1}/{\left(\alpha_\rL\alpha_{\rR} + \alpha_\rL-\alpha_\rR\right)}, 
$
$\omega_{\rm las}=c{m}/{\rho n_\text{L}}.
$
Finally, in the $\PT$-symmetric case $\alpha_{\rR}\alpha_\rL=1$,  CPA and lasing, of the second order, occur simultaneously, i.e., at the same value of $\tau^2$ [Fig.~\ref{scenarios01}(c)] but at different frequencies: $\omega_{\rm CPA}\ne \omega_{\rm las}$ (similarly  the  to CPA-and-laser discussed in~\cite{Chao} for simple SSs).
 
A peculiarity  of the described device is that spectral singularities of second (and higher) orders do occur for a set of incident wavelengths corresponding to the resonances in the $\mu$R. This can be useful for designing a laser or a CPA for frequency combs~\cite{comb}.  For example, for silica microrings ($n=1.45$) of radius $\rho=100~\mu$m (an $m$-th resonant mode has wavelength $\lambda_m=2\pi \rho/m$),
two neighboring resonances occur at wavelengths $\lambda_{700}\approx898~$nm and $\lambda_{699}\approx 899$~nm, 
which corresponds to the frequency interval $\Delta f\approx 0.329\,$THz between neighboring maxima of the comb.

\bigskip
The work of D.A.Z. is supported by the Russian Foundation for Basic Research (RFBR), Grant no. 19-02-00193$\backslash$19 and  by the government
of the Russian Federation (Grant No. 08-08).

\newpage

\section*{References with titles}


\begin{thebibliography}{99}
	
\bibitem{review1}  Y.\ Yamamoto and R.\ E.\ Slusher,   
Optical Processes in microcavities. 
Phys.\ Today {\bf 46}, 66 (1993).
	
\bibitem{Vahala} K. J.\ Vahala,
 Optical microcavities. 
Nature {\bf 424} 839-846 (2003).
	
\bibitem{review2} D. V. Strekalov, C. Marquardt, A. B. Matsko, H. G. L. Schwefel, and G. Leuchs,
		Nonlinear and quantum optics with whispering gallery resonators.
		J.\ Opt.\ {\bf 18}, 123002 (2016).

\bibitem{Yariv} A. Yariv, 
Universal relations for coupling of optical power between microresonators and dielectric waveguides.
Electron. Lett. {\bf 36}, 321 (2000). 

\bibitem{CaPaVa2000} M. Cai, O. Painter, and K. J.\ Vahala, 
Observation of Critical Coupling in a Fiber Taper to a Silica-Microsphere Whispering-Gallery Mode System,
Phys.\ Rev.\ Lett. {\bf 85}, 74 (2000)

\bibitem{Heebner}	J.\ E. Heebner,   R. W. Boyd, and Q-H. Park,	
SCISSOR solitons and other novel propagation effects in microresonator-modified waveguides.
J.\ Opt.\ Soc.\ Am.\ B {\bf 19}, 722 (2002).

\bibitem{Feng2014} \rtext{L. Feng,  Z. J. Wong,  R.-M. Ma, Y. Wang, and X. Zhang, 
Single-mode laser by parity-time symmetry breaking. 
Science {\bf 346}, 972–975 (2014)}.

\bibitem{Longhi} S. Longhi, and L. Feng,
 $\PT$ -symmetric microring laser-absorber.  
 Opt. Lett.\ {\bf 39}, 5026-5029 (2014).

\bibitem{three_cav} J.\ Li, R. Yu, C. Ding, and Y. Wu,
$\PT$ -symmetry-induced evolution of sharp asymmetric line shapes and high-sensitivity refractive
index sensors in a three-cavity array.
Phys.\ Rev.\ A {\bf 93}, 023814 (2016).
		
 

\bibitem{phot_molec}	
Y. Li,  F. Abolmaali, K. W. Allen,  N. I. Limberopoulos, A. Urbas, Y. Rakovich, A. V. Maslov, and V. N. Astratov, 
Whispering gallery mode hybridization in photonic molecules.
Las. Phot. Rev. {\bf 11},  1600278 (2017).

 \bibitem{SoCaAl} D. L. Sounas, C. Caloz, and A. Al\'u,
 Giant non-reciprocity at the subwavelength scale using angular momentum-biased metamaterials.
Nat. Comm. {\bf 4},  2407 (2013).

\bibitem{nonrecip1} B. Peng,  \c{S}. K. \"Ozdemir, F. Lei, F. Monifi, M. Gianfreda, G. L. Long,
S. Fan, F. Nori, C. M. Bender, and L. Yang,   Parity–time-symmetric whispering-gallery microcavities.
Nat. Phys. {\bf 10}, 394 (2014).

 \bibitem{nonrecip2}  X. Jiang, C. Yang, H. Wu, S. Hua, L. Chang, Y. Ding, Q. Hua, and M. Xiao, 
 On-Chip Optical Nonreciprocity Using an Active Microcavity.
 Sci. Rep.  {\bf 6}, 38972 (2016).

\bibitem{reverse} B. Peng, \c{S}. K. \"Ozdemir, S. Rotter,  H. Yilmaz,  M. Liertzer,  F. Monifi, C. M. Bender, F. Nori,  and L. Yang, 
Loss-induced suppression and revival of lasing.
Science {\bf 346}, 328 (2014). 

\bibitem{Mostafazadeh2013} A. Mostafazadeh, 
 \rtext{Optical spectral singularities as threshold resonances.}
 Phys.\ Rev.\ A \rtext{{\bf 83}, 045801 (2011)}
	
\bibitem{Stone} Y.\ D.\ Chong,  L. Ge, H. Cao,  and A. D. Stone,    
	Coherent perfect absorbers: Time-reversed lasers. 
	Phys.\ Rev.\ Lett. {\bf 105}, 053901 (2010).

	
 
	
\bibitem{Longhi-laser} S. Longhi,  
	$\PT$-symmetric laser absorber. 
	Phys.\ Rev.\ A, {\bf 82}, 031801 (2010).
	
\bibitem{ReGu} K. N. Reddy and S. D. Gupta, 
	Cavity-controlled spectral singularity. 
	Opt.\ Lett.\ {\bf 39} 4595--4598  (2014).
	
 \bibitem{Christod} \rtext{H. Hodaei, M.-A. Miri, M. Heinrich, D. N. Christodoulides, and M. Khajavikhan. 
 Parity-time–symmetric microring lasers.
 Science {\bf 346}, 975-978 (2014).}
	
 	
 \bibitem{doubleSS} V. V. Konotop,  E. Lakshtanov, and B. Vainberg, 
 Engineering coherent perfect absorption and lasing
 arXiv:1812.05499 [physics.optics]	
	  
	
 
\bibitem{KonZez2017} V. V. Konotop and D. A. Zezyulin,   
Phase transition through the splitting of self-dual spectral singularity in optical potentials. 
Opt.\ Lett.\ {\bf 42}, 5206 (2017).

\bibitem{Chao} C. Hang, G. Huang, and V. V. Konotop, 
Tunable spectral singularities: coherent perfect absorber and laser in an atomic medium. 
New J.\ Phys.\ {\bf 18}, 85003 (2016). 

\bibitem{comb} Th. Udem, R. Holzwarth and T. W. H\"ansch, 
	Optical frequency metrology.
Nature {\bf 416}, 233 (2002).	
	
	
 
\end{thebibliography}
\end{document}